\documentclass[final,3p,times,twocolumn]{elsarticle}

\usepackage{amsfonts}
\usepackage{mathrsfs}
\usepackage{comment}
\usepackage{xcolor}

\usepackage{graphicx}
\usepackage{amssymb}
\usepackage{amsmath}
\usepackage{amsthm}
\usepackage{lineno,hyperref}
\usepackage{color}

\journal{Journal of \LaTeX\ Templates}

\def\L{\mathscr{L}}

\begin{document}

\begin{frontmatter}

\title{Can electromagnetic charge inhabit in Rastall gravity?}

\author[1,2,3]{Bobir Toshmatov\corref{cor}}\cortext[cor]{Corresponding author}\ead{toshmatov@astrin.uz}
\author[2]{Zden\v{e}k Stuchl\'{i}k}\ead{zdenek.stuchlik@physics.slu.cz}

\author[3,4,5]{Bobomurat Ahmedov}\ead{ahmedov@astrin.uz}

\address[1]{New Uzbekistan University, Mustaqillik Ave. 54, Tashkent 100007, Uzbekistan}
\address[2]{Research Centre of Theoretical Physics and Astrophysics, Institute of Physics,\\ Silesian University in Opava, Bezru\v{c}ovo n\'{a}m. 13, CZ-74601 Opava, Czech Republic}
\address[3]{Ulugh Beg Astronomical Institute, Astronomy 33, Tashkent 100052, Uzbekistan}

\address[4]{IFAR, ``TIIAME" NRU, Kori Niyoziy 39, Tashkent 100000, Uzbekistan}
\address[5]{National University of Uzbekistan, Tashkent 100174, Uzbekistan}

\begin{abstract}
One of the eminent generalizations of theory of general relativity is the Rastall gravity which was {constructed} based on the assumption of the non-conserved energy-momentum tensor of the matter field. Despite in the literature several solutions of black holes in the Rastall gravity coupled to the electromagnetic field have been presented, in the current paper we argue that the Rastall gravity with non-conserved energy-momentum tensor (with $\lambda\neq0$ and $R\neq0$) cannot couple to the electrodynamics, i.e., the electromagnetically charged black hole solution cannot be obtained in this case. This statement is adequate for both linear and nonlinear electrodynamics with the electric, magnetic, or dyonic charges coupled to the Rastall gravity.
\end{abstract}

\begin{keyword}
General relativity \sep Rastall gravity \sep electrodynamics 
\end{keyword}

\end{frontmatter}

\section{Introduction}\label{sec-intro}

General relativity is the prevailing theory of gravity, which describes the force as a result of massive objects warping spacetime. It is well-known that the theory of general relativity is one of the simplest and most elegant theories of gravitation that has explained several hidden mysteries of the nature and universe. The correctness of general relativity has been verified by several experimental tests in the weak field regimes \cite{Will:2014kxa}. However, its correctness in the strong gravitational fields is still under debate, as recent observational breakthrough events such as detection of the gravitational waves from the coalescence of two black holes or neutron stars in binaries \cite{GW151226,GW170104,GW170814,GW170608,GW170817} and obtained images of the supermassive black holes in the centers of Milky Way \cite{EHTMW} and M87 \cite{EHT} galaxies have still left some room for the alternative and modified theories of gravity. Furthermore, there are still big numbers of shortcomings in theoretical and observational aspects of the science, such as accelerated universe, the existence of dark matter and energy, on account of which modification of general relativity is inevitable.

There are possible modified and alternative theories of gravity that have been constructed in mainly two ways. In the first method, the fundamental assumptions of general relativity are still kept but new additional terms are added to the Lagrangian density of the general relativity on account of which all field equations are modified with additional terms. In the second method, the fundamental assumptions of general relativity are changed. The Rastall gravity belongs to the latter group. It is clear that one of the fundamental assumptions of general relativity is the null covariant divergence of the energy-momentum tensor. However, Rastall followed the second method and attempted to modify the theory of general relativity by assuming that the covariant derivative of the energy-momentum tensor varies directly as the derivative of the Ricci scalar \cite{Rastall:1972swe,Rastall:1976uh}. Clearly, it explicitly represents a nonminimal coupling between matter and geometry in curved spacetimes. {In flat spacetimes, the energy-momentum tensor satisfies a conservation law as the curvature of spacetime vanishes. As a result, the field equations become equivalent to the vacuum Einstein equations. Even in the curved spacetime with matter field, the field equations of the Rastall gravity can be as the Einstein equations in terms of the effective energy-momentum tensor which we will present later in the next section. From this point of view, The Rastall gravity can be considered as the equivalent to general relativity as stated in \cite{Visser:2017gpz}. Moreover, these theories are equivalent in vacuum and in non-vacuum case with conserved energy-momentum tensors. However, in the matter field with non-conserved energy-momentum tensor, these two theories are different.} In spite of that, there are still various issues that these two theories result from different consequences \cite{Darabi:2017coc}. Over the last decades, Rastall gravity has gained attention of scientists and as a result of it, a vast number of papers related to the new solutions, or properties of the solutions of black holes in Rastall gravity, have been published \cite{Batista:2011nu,Oliveira:2015lka,Ghosh:2021byh,Maulana:PRD:2019,ElHanafy:2022kjl}. Among them there are the ones related to the charged solutions \cite{Sakti:2019krw,Sakti:2021gru,Guo:2021bwr,Nashed:2022dkj,Gogoi:2021dkr,Shao:2022oqv,Narzilloev:2022bbs}. Unsurprisingly, in this paper we present results that question possibility of obtaining charged black hole solutions in Rastall gravity.

We study the possibility of the construction of the black hole solution in Rastall gravity coupled to the electrodynamics. The paper is organized as follows: In section \ref{sec-background} we revisit Rastall gravity and equations of motion. Section \ref{sec-Maxwell} is devoted to a system of Rastall gravity coupled to the Maxwell electrodynamics. In section \ref{sec-ned} we repeat the calculations presented in the previous section for the nonlinear electrodynamics. Finally, in sections \ref{sec-discussion} and \ref{sec-conclusion} we discuss and summarize the main results obtained in the paper. Throughout the paper, we use the geometrized units in which the Newtonian gravitational constant $G_N$, and speed of light $c$ are set as $G_N=c=1$.

\section{Rastall gravity revisited}\label{sec-background}

\begin{figure}
    \centering
    \includegraphics[scale=0.45]{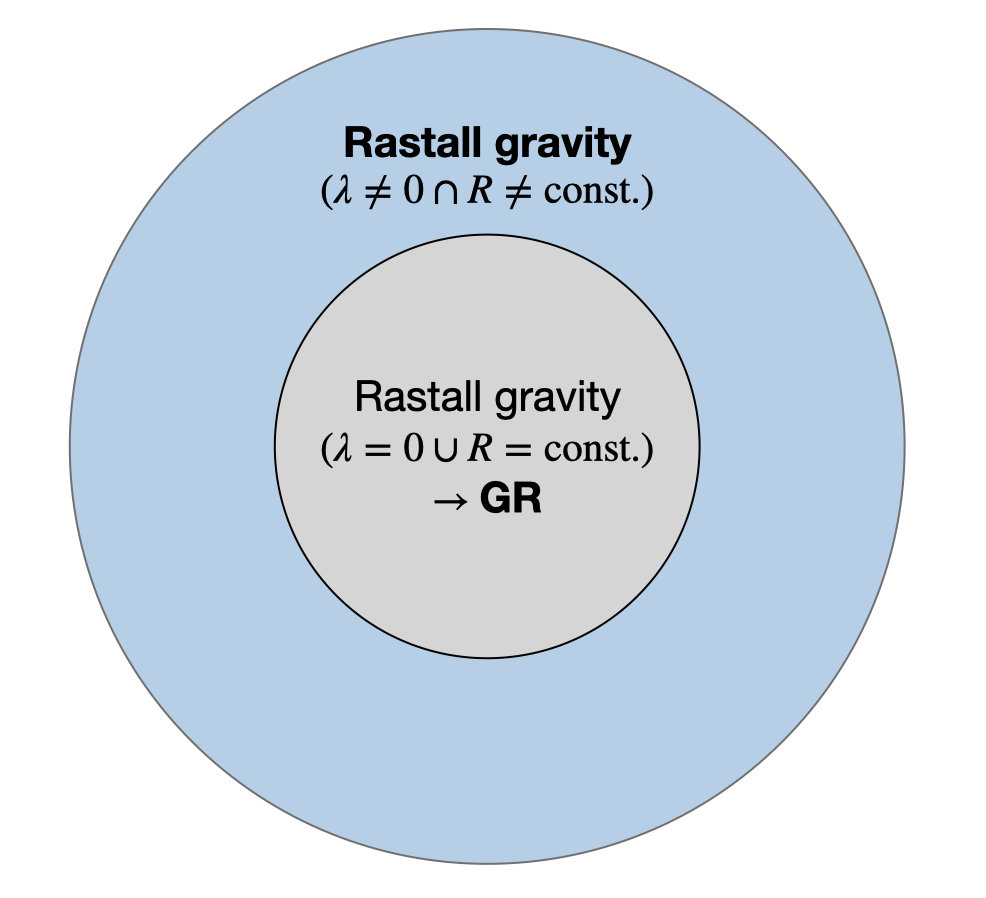}
    \caption{Schematic picture of the Rastall gravity that generalizes general relativity (GR) is presented.}
    \label{fig-schematic}
\end{figure}

According to Rastall's theory of gravity, the conservation of the energy-momentum tensor, $T^{\mu\nu}$, is assumed to be given by \cite{Rastall:1972swe}
\begin{eqnarray}\label{rastall}
   \nabla_\nu T^{\mu\nu}=\lambda \nabla^\mu R\ ,
\end{eqnarray}
where $\lambda$ is a free parameter, so-called Rastall coupling parameter and $R$ is a Ricci scalar. The nondivergence-free energy-momentum assumption that varies directly as the derivative of Ricci scalar, represents a nonminimal coupling between matter and geometry. In other words, if there is no matter field (vacuum), the theory of general relativity is recovered. Thus, Rastall gravity being generalization of general relativity, it includes general relativity in which the energy-momentum tensor conserved that corresponds to $\lambda=0$ or $R={\rm const.}$ (inner region in Fig. \ref{fig-schematic}) and the case in which the energy-momentum tensor is not conserved that corresponds to $\lambda\neq0$ and $R\neq {\rm const.}$ (outer region in Fig. \ref{fig-schematic}).

According to the assumption of a non-divergence-free energy-momentum tensor in the curved spacetime, Rastall obtained a consistent set of field equations
\begin{eqnarray}\label{field-eq}
    \tilde{G}_{\mu\nu}\equiv G_{\mu\nu}+\lambda \kappa g_{\mu\nu}R=\kappa T_{\mu\nu}\ ,
\end{eqnarray}
with the Einstein tensor defined by
\begin{eqnarray}
   G_{\mu\nu}=R_{\mu\nu}-\frac{1}{2}g_{\mu\nu}R\ ,
\end{eqnarray}
where $\kappa=8\pi$. In the case of general relativity which corresponds to $\lambda=0$, equation (\ref{field-eq}) reduces to the well-known Einstein's field equations $G_{\mu\nu}=\kappa T_{\mu\nu}$. The trace of the field equation (\ref{field-eq}) results the following equation:
\begin{eqnarray}\label{trace-R}
    R=\frac{\kappa}{4\kappa\lambda-1}T\ , \qquad \kappa\lambda\neq\frac{1}{4}\ .
\end{eqnarray}
In terms of the expression (\ref{trace-R}), one can rewrite the field equation (\ref{field-eq}) in the alternative form as
\begin{eqnarray}\label{equiv}
    G_{\mu\nu}=\kappa \tilde{T}_{\mu\nu}\ ,
\end{eqnarray}
where 
\begin{eqnarray}
    \tilde{T}_{\mu\nu}=T_{\mu\nu}-\frac{\kappa\lambda}{4\kappa\lambda-1}g_{\mu\nu}T\ .
\end{eqnarray}
One can see from (\ref{equiv}) that the Rastall term in the field equation can be considered as a new additional matter field to the right-hand side of the Einstein equations. Therefore, in this sense, Rastall gravity can be considered formally equivalent to general relativity \cite{Visser:2017gpz}. Now by choosing the appropriate matter field and energy-momentum tensor and fixing the background ansatz, one can construct the solution in Rastall gravity. For simplicity, we choose the static, spherically symmetric spacetime as
\begin{eqnarray}\label{metric-ansatz}
    ds^2=-f(r)dt^2+\frac{dr^2}{f(r)}+r^2\left(d\theta^2+\sin^2\theta d\phi^2\right)\ . 
\end{eqnarray}
For the spacetime line element (\ref{metric-ansatz}) non-zero components of the left hand side of the field equations (\ref{field-eq}) are given as
\begin{eqnarray}\label{einstein-tensor}
    &&\tilde{G}^t_{\ t}=\frac{r f'+f-1}{r^2}+\lambda\kappa R=\tilde{G}^r_{\ r}\ ,\nonumber\\
    &&\tilde{G}^\theta_{\ \theta}=\frac{f''}{2}+\frac{f'}{r}+\lambda\kappa R=\tilde{G}^\phi_{\ \phi}\ ,
\end{eqnarray}
where the Ricci scalar equals
\begin{eqnarray}\label{ricci}
    R=-f''-\frac{2\left(2 r f'+f-1\right)}{r^2}\ .
\end{eqnarray}
In the following section, we obtain such solutions in Rastall gravity for electrically charged matter fields. The solutions will be obtained through mathematical calculations and analysis in the framework of the newly developed formalism. 

\section{Maxwell electrodynamics coupled to Rastall gravity}\label{sec-Maxwell}

Since in the previous section all equations related to the geometry of the spacetime are given, in this section we solve these equations by setting the energy-momentum tensor of the matter field by the Maxwell electrodynamics:  
\begin{eqnarray}\label{em-tensor-Maxwell}
T_{\mu\nu}=\frac{2}{\kappa}\left(F_\mu^{\ \alpha} F_{\nu\alpha}- \frac{1}{4}g_{\mu\nu} F_{\delta\gamma} F^{\delta\gamma}\right)\ ,
\end{eqnarray}
where the electromagnetic field tensor is determined by the vector potential $A_\mu$ as $F_{\mu\nu}=\partial_{\mu}A_\nu-\partial_\nu A_\mu$. Their explicit forms are identical for both electrically and magnetically charged cases as
\begin{eqnarray}\label{emt-Maxwell}
    T^t_{\ t}=T^r_{\ r}=-\frac{Q^2}{\kappa r^4}\ ,\qquad T^\theta_{\ \theta}=T^\phi_{\ \phi}=\frac{Q^2}{\kappa r^4}\ .
\end{eqnarray}
By simplifying the field equations (\ref{equiv}) with (\ref{einstein-tensor}) and (\ref{emt-Maxwell}), we obtain the following two independent differential equations:
\begin{eqnarray}\label{dif-eq1}
\frac{1-r f'-f}{r^2}-\lambda\kappa R=\frac{Q^2}{r^4},
\end{eqnarray} 
\begin{eqnarray}\label{dif-eq2}
\frac{f''}{2}+\frac{f'}{r}+\lambda\kappa R=\frac{Q^2}{r^4}\ .
\end{eqnarray}
If the above differential equations are solved, separately, in general relativity limit ($\lambda=0$), both equations (\ref{dif-eq1}) and (\ref{dif-eq2}) would result in the well-known Reissner-Nordstr\"{o}m solution
\begin{eqnarray}\label{rn-metric}
    f(r)=1-\frac{2M}{r}+\frac{Q_e^2}{r^2}\ .
\end{eqnarray}
From the fact that the Reissner-Nordstr\"{o}m solution (\ref{rn-metric}) is obtained by solving both differential equations (\ref{dif-eq1}) and (\ref{dif-eq2}), we are convinced that the linear electrodynamics can couple to the general relativity.

In the next step, we solve the differential equations (\ref{dif-eq1}) and (\ref{dif-eq2}) separately in Rastall gravity ($\lambda\neq0$). By solving the first differential equation (\ref{dif-eq1}), we obtain the metric function as following:
\begin{eqnarray}\label{sol1}
    f(r)=1-\frac{2M}{r}+\frac{Q^2}{r^2}+c r^{-2+\frac{1}{\lambda\kappa}}\ ,
\end{eqnarray}
while from the second differential equation (\ref{dif-eq2}) we arrive at the metric function 
\begin{eqnarray}\label{sol2}
    f(r)=1-\frac{2M}{r}+\frac{Q^2}{r^2}+c r^{\frac{4\lambda\kappa}{1-2\lambda\kappa}}\ ,
\end{eqnarray}
where $c$ is the integration constant. For $f(r)$ to be the solution, both metric functions (\ref{sol1}) and (\ref{sol2}) must be identical. This can happen only in the following two cases:
\begin{itemize}
    \item[(i)] the integration constant is zero ($c=0$). In this case both solutions are reduced to the Reissner-Nordstr\"{o}m black hole solution. We should note that the Reissner-Nordstr\"{o}m solution is also the solution of the Rastall gravity only because of the fact that it is a solution of general relativity coupled to the Maxwell electrodynamics. We strengthen our this statement with the fact that for the Reissner-Nordstr\"{o}m solution the Ricci scalar vanishes ($R=0$) and consequently, energy-momentum tensor becomes conserved and all field equations (including (\ref{rastall}), (\ref{field-eq}) and (\ref{equiv})) take the form of the ones of pure general relativity;
    \item[(ii)] exponents of $r$ in the last terms of both metric functions (\ref{sol1}) and (\ref{sol2}) are equal when $\lambda\kappa=1/4$. In this case the metric function takes the de-Sitter-like form as (it was obtained in \cite{Nashed:2022dkj})
\begin{eqnarray}\label{not-sol}
    f(r)=1-\frac{2M}{r}+\frac{Q^2}{r^2}+c r^2\ .
\end{eqnarray}
However, even if the metric function (\ref{not-sol}) satisfies both differential equations (\ref{dif-eq1}) and (\ref{dif-eq2}), it does not satisfy the trace of the field equation (\ref{trace-R}), as $\lambda\kappa=1/4$ is the forbidden value (see also \cite{Rastall:1972swe}). This means that even though the metric function satisfies both differential equations, it fails to satisfy a crucial condition that needs to be met in order to be considered a valid solution. The restricted value of $\lambda\kappa=1/4$ is an essential requirement that the metric function must fulfill in order to be considered a solution to the field equations. Without satisfying this condition, it can be concluded that the metric function is not a valid solution. Therefore, it is not a solution. 
\end{itemize}
Thus, as we have discussed in above two cases that the charged black hole solution can exist only in general relativity that is a subset of Rastall gravity.

\section{Nonlinear electrodynamics coupled to Rastall gravity}\label{sec-ned}

We have shown in the previous section that Maxwell electrodynamics cannot couple to Rastall gravity to produce charged black hole solution. As a continuation, we generalize the scenario to the nonlinear electrodynamics coupled to Rastall gravity. We again adopt the spacetime in a static, spherically symmetric form. The energy-momentum tensor of the nonlinear electrodynamics is given by
\begin{eqnarray}\label{em-tensor}
T_{\mu\nu}=\frac{2}{\kappa}\left(\L_FF_\mu^{\ \alpha} F_{\nu\alpha}- \frac{1}{4}g_{\mu\nu}\L\right)\ ,
\end{eqnarray}
where $\L_F$ denotes a derivative of the Lagrangian density of the nonlinear electrodynamics with respect to invariant, $F\equiv F_{\mu\nu}F^{\mu\nu}$, $\L_F=\partial_F\L$. In general, the electromagnetic four-potential, $A_\mu$, is written for the spherically symmetric spacetimes in terms of the 1-forms of Carter tetrad for the background line element (\ref{metric-ansatz}) as \cite{Znajek:MNRAS:1977}
\begin{eqnarray}\label{4-pot}
    A=-\frac{\varphi}{\sqrt{f}}\omega^t-\frac{Q_m\cot\theta}{r}\omega^\phi\ .
\end{eqnarray}
where the 1-forms of the Carter tetrad is defined as
\begin{eqnarray}\label{tetrads}
    &&\omega^t=\sqrt{f}dt\ , \qquad \omega^r=\frac{1}{\sqrt{f}}dr\ ,\nonumber\\
    &&\omega^\theta=rd\theta\ , \qquad \omega^\phi=r\sin\theta d\phi\ .
\end{eqnarray}
Thus, the electromagnetic field four-potential (\ref{4-pot}) can be written explicitly as
\begin{eqnarray}\label{4-potential}
    A=\varphi(r)dt-Q_m\cos\theta d\phi\ ,
\end{eqnarray}
where the time and azimuthal components of the potential (\ref{4-potential}) correspond to the electrically and magnetically charged black holes, respectively, as $\varphi$ and $Q_m$ are electric scalar potential and magnetic charge, respectively. Unlike the case of linear electrodynamics in a previous section, in the case of nonlinear electrodynamics electrically and magnetically charged electromagnetic fields do not generate the identical solutions \cite{Bronnikov:PRL:2000,Bronnikov:PRD:2001,Fan:PRD:2016,Bronnikov17,Toshmatov:PRD:comment} and their characteristic oscillations are not isospectral \cite{Toshmatov:2018tyo,Toshmatov:2021fgm,Moreno:2002gg,Nomura:2020tpc,Nomura:2021efi}. Therefore, below we present a construction of the electrically and magnetically charged black hole solutions in Rastall gravity coupled to the nonlinear electrodynamics, separately.

\subsection{Electrically charged black holes}

In the case of the electrically charged black hole, the time component of the four-potential (\ref{4-potential}) of the electromagnetic field survives as
\begin{eqnarray}
    A_\mu=\{\varphi(r),\ 0,\ 0,\ 0\}\ .
\end{eqnarray}
After calculating the components of the electromagnetic field tensor, we find the electromagnetic field invariant as
\begin{eqnarray}\label{invariant-el}
    F_e=-2\varphi'^2\ .
\end{eqnarray}
The nonvanishing components of the energy-momentum tensor of the nonlinear electrodynamics are found to be
\begin{eqnarray}\label{emt-electric}
T^t_{\ t}&=&-\frac{\L-2\L_{F}F_e}{\kappa}=T^r_{\ r}\ ,\nonumber\\ 
T^\theta_{\ \theta}&=&-\frac{\L}{\kappa}=T^\phi_{\ \phi}\ . 
\end{eqnarray}
By solving the field equations  (\ref{field-eq}) with (\ref{einstein-tensor}) and (\ref{emt-electric}), we obtain the following expressions for the Lagrangian density of the nonlinear electrodynamics:
\begin{eqnarray}
\L&=&-f''-\frac{2f'}{r}-2\lambda\kappa R,\label{lagrangian-electric1}
\\
\L_{F}&=&-\frac{r^2 f''-2 f+2}{2 r^2 F_e}\ .\label{lagrangian-electric2}
\end{eqnarray}
It can be easily noticed from the expressions (\ref{lagrangian-electric1}) and (\ref{lagrangian-electric2}) that the Lagrangian density of the nonlinear electrodynamics is the function of Rastall coupling parameter ($\partial\L/\partial\lambda\neq0$), while its derivative with respect to the electromagnetic invariant, $\L_F$, is independent of the Rastall coupling parameter ($\partial\L_F/\partial\lambda=0$). At this point, we come across a problem with the equations, as expressions of the Lagrangian density (\ref{lagrangian-electric1}) and (\ref{lagrangian-electric2}) contradict each other. Mathematically, if the Lagrangian density is the function of the parameter $\lambda$, its derivative with respect to the electromagnetic field invariant, $F_e$, which is not a function of the parameter $\lambda$ ($\partial F/\partial\lambda=0$) must also be the function of $\lambda$, as $\L_F=\L'/F'$ with the prime being the derivative with respect to the radial coordinate, $r$. However, it is not the case here. In short, if the differential equations (\ref{lagrangian-electric1}) and (\ref{lagrangian-electric2}) are solved for any Lagrangian density of the nonlinear electrodynamics separately, one would obtain the metric function with the Rastall term in the first solution (solution of (\ref{lagrangian-electric1})), while in the second one (solution of (\ref{lagrangian-electric2})) there is no term with the Rastall parameter. In other words, if one solves the differential equations for the nonlinear electrodynamics Lagrangian density separately, the first solution (the solution of (\ref{lagrangian-electric1})) would include the Rastall term, while the second solution (the solution of (\ref{lagrangian-electric2})) would not include it. Therefore, there are two cases to address this contradiction: 
\begin{itemize}
    \item[(i)] $\lambda=0$ that corresponds to general relativity coupled to the nonlinear electrodynamics, as the energy-momentum tensor  becomes conserved. The obtained solution can be considered the solution of the Rastall gravity only because of the fact that general relativity is a subset of the Rastall gravity;
    \item[(ii)] $R=0$ that corresponds to general relativity coupled to the nonlinear electrodynamics. However, since $R=0$ implies the solution to be Reissner-Nordstr\"{o}m-like and it is the solution of general relativity coupled to the linear electrodynamics, this system does not provide any solution not only in the general relativity coupled to the nonlinear electrodynamics, but also in Rastall gravity coupled to the nonlinear electrodynamics. 
\end{itemize}
Thus, in both cases the final solution would be the one in general relativity and not in Rastall gravity with nonconserved energy-momentum tensor. Thus, we are convinced that one cannot obtain the electrically charged black hole solution in Rastall gravity with $\lambda\neq0$ and $R\neq0$ coupled to the nonlinear electrodynamics.

\subsection{Magnetically charged black holes}

In the magnetically charged black hole solution, the four potential of the electromagnetic field (\ref{4-potential}) is given by
\begin{eqnarray}
    A_\mu=\{0, 0, 0, -Q_m\cos\theta\}\ .
\end{eqnarray}
In this case, the electromagnetic field invariant takes the following form:
\begin{eqnarray}\label{invariant-mag}
    F_m=\frac{2Q_m^2}{r^4}\ .
\end{eqnarray}
The independent components of the energy-momentum tensor of the nonlinear electrodynamics are given as
\begin{eqnarray}\label{emt-magnetic}
    T^t_{\ t}&=&-\frac{\L}{\kappa}=T^r_{\ r}\ ,\nonumber\\ 
    T^\theta_{\ \theta}&=&-\frac{\L-\L_{F}F_m}{\kappa}=T^\phi_{\ \phi}\ . 
\end{eqnarray}
By solving the field equations  (\ref{field-eq}) with (\ref{einstein-tensor}) and (\ref{emt-magnetic}), we obtain the following expressions for the Lagrangian density:
\begin{eqnarray}
    \L&=&\frac{2\left(1-r f'-f\right)}{r^2}-2\lambda\kappa  R,\label{lagrangian-magnetic1}
    \\
    \L_F&=&\frac{r^2 f''-2 f+2}{2r^2F_m}\ .\label{lagrangian-magnetic2}
\end{eqnarray}
Thus, the situation is the same as in the case of the electrically charged black hole solution in a previous subsection, and consequently the conclusion is also similar. Since the results in both electrically and magnetically charged cases are the same, it is obvious that it remains unchanged in the dyonically charged case too. Therefore, we do not here report this case.

\section{Discussion}\label{sec-discussion}

In previous sections, we have shown by solving the field equations  (\ref{field-eq}) that the electromagnetic field cannot couple to Rastall gravity in black hole solutions. However, in that process, we have not used the equation on non-conserved energy-momentum tensor in Rastall gravity (\ref{rastall}) and the Maxwell equations for the nonlinear electrodynamics which are given by
\begin{eqnarray}\label{Maxwell-eq}
    \nabla_\mu\left(\L_F F^{\mu\nu}\right)=0\ .
\end{eqnarray}
These two equations contradict each other, as equation (\ref{Maxwell-eq}) requires the covariant derivative of the right hand side of the field equation (\ref{field-eq}) to vanish, while the covariant derivative of the left hand side of (\ref{field-eq}) is nonzero via (\ref{rastall}). To overcome this problem, we must consider either $\lambda=0$ or $R=0$ cases. The former case represents the absence of the Rastall term in the equations of motion, i.e., it corresponds to general relativity with Einstein equations coupled to the nonlinear electrodynamics. This type of solutions are obtained and well studied in the vast literature such as \cite{Bronnikov:PRL:2000,Bronnikov:PRD:2001,Fan:PRD:2016}. In the latter case when the Ricci scalar (or trace of the energy-momentum tensor) vanishes, there is only one solution that corresponds to the Reissner-Nordstr\"{o}m-like one as
\begin{eqnarray}\label{ricci0}
    R=0 \quad \Rightarrow \quad f(r)=1-\frac{2M}{r}+\frac{c}{r^2}\ .
\end{eqnarray}
with $c$ being the integration constant. From this solution it turns out that to have all field equations satisfied, the following equation also must give the same metric function as (\ref{ricci0}):
\begin{eqnarray}\label{field-eq-ricci0}
    R_{\mu\nu}=\kappa T_{\mu\nu}\ .
\end{eqnarray}
Despite the field equations have changed, they still do not represent the general Rastall gravity containing nonzero $\lambda$. The solution of the field equation (\ref{field-eq-ricci0}) for the Maxwell electrodynamics represents the Reissner-Nordstr\"{o}m black hole solution which is again independent from the Rastall parameter.

By solving equations (\ref{field-eq-ricci0}) for the nonlinear electrodynamics (\ref{em-tensor}) with metric function (\ref{ricci0}), we obtain that the Lagrangian density of the nonlinear electrodynamics is linearly related to the electromagnetic field invariant, $F$, as
\begin{eqnarray}
    \L\propto F, \qquad \L_F\propto {\rm constant}\ ,
\end{eqnarray}
whose solution is independent from the Rastall parameter. From this approach also we have confirmed that Rastall gravity (with $\lambda\neq0$ and $R\neq0$) cannot coexist with the electrodynamics.

\section{Conclusion}\label{sec-conclusion}

Rastall gravity is a simple mathematical generalization of general relativity on account of the non-minimal coupling between the geometry and energy-momentum tensor of the matter field. The original physical motivation of this theory was that the local energy-momentum conservation law in the flat spacetime does not necessarily imply its conservation in the curved spacetime with matter field. This on turn implies several new interesting properties in the gravitational theory. One of such properties has been addressed in this paper. Despite in many literature new black hole solutions have been obtain within the framework of Rastall gravity coupled to the electrodynamics (sometimes some additional second matter field is also considered), we have shown here that in this system one cannot obtain the electromagnetically (electrically, magnetically, or dyonically) charged black hole solutions in Rastall gravity with non-conserved energy-momentum tensor ($\lambda\neq0$ and $R\neq0$) coupled not only to the linear but also to the nonlinear electrodynamics. In other words, the electromagnetically charged solutions of black hole cannot be obtained in Rastall gravity corresponding to the outer region of Fig. \ref{fig-schematic}.

\section*{Acknowledgement}

BT acknowledges the support of program "CZ.02.2.69/0.0/0.0/18-053/0017871: Podpora mezin\'{a}rodn\'{i} mobility na Slezsk\'{e} univerzit\v{e} v Opav\v{e}" at the Institute of Physics, Silesian University in Opava. Authors acknowledge the support of Ministry of Innovative Development of the Republic of Uzbekistan Grants No. F-FA-2021-432 and MRB-2021-527.

\bibliographystyle{iopart}
\bibliography{references}
\end{document}